\documentclass[11pt]{article}
\usepackage{amsmath,amssymb,color,graphics,epsfig,cite}

\usepackage{amsfonts}

\newcommand{\w}[1]{\\[0.#1cm]}

\newcommand{\be}{\begin{equation}}
\newcommand{\ee}{\end{equation}}
\newcommand{\bea}{\setlength\arraycolsep{2pt} \begin{eqnarray}}
\newcommand{\eea}{\end{eqnarray}}
\newcommand{\nn}{\nonumber}

\def\ft#1#2{{\textstyle{\frac{\scriptstyle #1}{\scriptstyle #2} } }}

\def\0{{\sst{(0)}}}
\def\1{{\sst{(1)}}}
\def\2{{\sst{(2)}}}
\def\3{{\sst{(3)}}}
\def\4{{\sst{(4)}}}
\def\5{{\sst{(5)}}}
\def\6{{\sst{(6)}}}
\def\7{{\sst{(7)}}}
\def\8{{\sst{(8)}}}
\def\sst#1{{\scriptscriptstyle #1}}

\def\a{\alpha}
\def\b{\beta}

\def\G{\Gamma}
\def\d{\delta}

\def\l{\lambda}
\def\L{\Lambda}
\def\m{\mu}
\def\n{\nu}
\def\r{\rho}
\def\s{\sigma}
\def\S{\Sigma}

\def\th{\theta}
\def\Th{\Theta}

\def\O{\Omega}

\begin{document}

\begin{center}
{\Large {\bf Improved Wald Formalism and First Law of Dyonic Black Strings with Mixed Chern-Simons Terms }}

\vspace{20pt}
{\large Liang Ma, Yi Pang, H. L\"u}

\vspace{10pt}

{\it Center for Joint Quantum Studies and Department of Physics\\ School of Science, Tianjin University,\\ Yaguan Road 135, Jinnan District, Tianjin 300350, China}
\vspace{40pt}

\underline{ABSTRACT}
\end{center}

We study the first law of thermodynamics of dyonic black strings carrying a linear momentum
in type IIA string theory compactified on K3 with leading order $\alpha'$ corrections. The low energy effective action contains mixed Chern-Simons terms of the form $-2B_{\2}\wedge {\rm tr}(R(\G_\pm)\wedge R(\G_\pm))$ which is equivalent to $2H_{\3}\wedge \mathrm{CS}_{\3}(\G_\pm)$ up to a total derivative. We find that the naive application of Wald entropy formula leads to two different answers associated with the two formulations of the mixed Chern-Simons terms. Surprisingly, neither of them satisfies the first law of thermodynamics for other conserved charges computed unambiguously using the standard methods. We resolve this problem by carefully evaluating the full infinitesimal Hamiltonian at both infinity and horizon, including
contributions from terms proportional to the Killing vector which turn out to be nonvanishing on the horizon and indispensable to establish the first law.  We find that the infinitesimal Hamiltionian associated with $-2B_{\2}\wedge {\rm tr}(R(\G_\pm)\wedge R(\G_\pm))$ requires an improvement via adding a closed but non-exact term, which vanishes when the string does not carry either the magnetic charge or linear momentum. Consequently, both formulations of the mixed Chern-Simons terms yield the same result of the entropy that however does not agree with the Wald entropy formula. In the case of extremal black strings, we also contrast our result with the one obtained from Sen's approach.

\vfill{\footnotesize  liangma@tju.edu.cn\ \ \ pangyi1@tju.edu.cn\ \ \ mrhonglu@gmail.com}



\thispagestyle{empty}
\pagebreak

\tableofcontents
\addtocontents{toc}{\protect\setcounter{tocdepth}{2}}


\section{Introduction}

Since the establishment of black hole mechanics \cite{Bardeen:1973gs} in 1973,  many methods have been proposed to compute the thermodynamic quantities, amongst which the most notable ones include the Euclidean action method \cite{Gibbons:1976ue}, the construction of the quasi-local conserved charges \cite{Brown:1992br, Wang:2008jy} based on the ADM formalism \cite{Arnowitt:1959ah} and the ADT method \cite{Abbott:1981ff}. In these methods, the satisfaction of the first law of mechanics are verified independently after deriving the thermodynamical quantities. It was Wald \cite{Wald:1993nt,Iyer:1994ys} who first realized that the thermodynamical quantities and the first law of mechanics can be combined into one formula, {\it i.e.}, through the first law of mechanics, one can identify various thermodynamical quantities. By this way, the first law of mechanics is obeyed automatically. The idea is that given a Killing vector $\xi$ in $D$-dimensional spacetime, one can construct a closed $(D-2)$-form which locally can be written as \cite{Wald:1993nt,Iyer:1994ys,Barnich:2001jy}
\be
d\big(\delta\mathbf{Q}[\xi]-i_{\xi}\mathbf{\Theta}[\delta\phi]\big)=0\,,
\label{d-2form}
\ee
where $\delta\phi$ denotes variations of all the fields that satisfy the linearized field equations. Integration of the quantity inside the bracket on the $(D-2)$-dimensional hypersurface defines the infinitesimal Hamiltonian associated with the Killing vector $\xi$, {\it i.e.},
\be\label{integralform}
\delta{\cal H}_{\Sigma}=\int_{\Sigma}( \delta\mathbf{Q}[\xi]-i_{\xi}\mathbf{\Theta}[\d\phi])\,.
\ee
Applying this formalism to black holes, one evaluates \eqref{d-2form} on the constant time slice sandwiched between the
spatial infinity and the bifurcation horizon $\mathcal{B}$ to obtain
\be
\delta{\cal H}_{\infty}=\delta{\cal H}_{\mathcal{B}}.
\label{firstlaw}
\ee
Upon substituting a specific black hole solution, one recognizes the equality above gives precisely the first law of mechanics while the integral at infinity yields combinations of conserved charges such as mass, angular momentum; the integral on the horizon is used to
define the entropy \cite{Wald:1993nt,Iyer:1994ys}.

However, it was already pointed out in \cite{Iyer:1994ys} that the density of the infinitesimal Hamiltionian was defined up to an addition of a closed $(D-2)$-form. In fact, $\delta\mathbf{Q}[\xi]-i_{\xi}\mathbf{\Theta}[\d\phi]$ should be classified by the $(D-2)$'th cohomology class on the spacetime with a possible gauge bundle structure. In order for \eqref{firstlaw} to hold,
one must chose properly the density of the infinitesimal Hamiltionian such that it is globally well defined. A specific example is given by the Reissner-Nordstr\"om (RN) dyonic black hole in the $D=4$ Einstein-Maxwell theory
\be
e^{-1}{\cal L}_{\rm{EM}}=R-\frac{1}{4}F_\2^2\,,
\ee
where $F_\2= dA_\1$. A direct application of the Wald procedure leads to \cite{Lu:2013ura}
\bea
\mathbf{Q}[\xi]&=& -\star d\xi-(i_\xi A_\1)\star F_\2\,,\qquad i_\xi\mathbf{\Theta}=\frac12\epsilon_{\a\b\m\n}\Th^\a\xi^\b dx^\m\wedge dx^\n\,\nn\\
\Th^\a&=&g^{\a\sigma}g^{\nu \rho}\left(\nabla_\rho\delta g_{\nu\sigma}-\nabla_\sigma\delta g_{\nu\rho}\right)-F^{\a\rho}\delta A_\rho\,,
\label{EMexample}
\eea
where the star denotes the Hodge dual and $\xi$ is the Killing vector vanishing on the bifurcation horizon. In the gauge, $i_\xi A_{\1}|_{r=r_h}=0$, one finds the integral
at infinity yields $d M-\Phi_\mathrm{e}d Q_\mathrm{e}$ while the integral at the horizon gives $Td S$. This means the equality \eqref{firstlaw} is not satisfied, since the magnetic contribution is absent. To cure this problem,  one has to add a closed form $-d(\Psi\delta A_{\1})$ to $i_\xi\mathbf{\Theta}$ where $\Psi$ is defined by
\bea
d\Psi=i_\xi\star F_\2.
\eea
Then the newly defined $i_\xi\mathbf{\Theta}$ would contain a term $\Psi\delta F_{\2}$ whose value on the horizon provides the missing magnetic contribution $\Phi_\mathrm{m}d Q_\mathrm{m}$ to the first law \cite{Lu:2013ura}. Adding to the original infinitesimal Hamiltonian by a closed form $-d(\Psi\delta A_{\1})$ is also a requirement from regularity. Without such an improvement, it would contain a term proportional $\delta P dr\wedge \cos\th d\phi$ suffering from the Dirac string singularity, when the solution carries the magnetic charge.
The magnetic part in the first law can also be introduced via the electromagnetic duality \cite{Gibbons:2013dna} or a careful analysis based on the Hamiltonian formulation \cite{Copsey:2005se}. The trick of pulling out a total derivative was also needed in the generalization of the original proof of the first law to Einstein gravity coupled to a non-linear electrodynamic system \cite{Rasheed:1997ns}.

In order to see that the closed form $-d(\Psi\delta A_{\1})$ naturally comes from electromagnetic duality, one can simply repeat the
Wald procedure for the dual Lagrangian of Einstein-Maxwell theory ${\cal L }(g,\widetilde{A})$ where $d\widetilde{A}_{(1)}=\star F_{(2)}$, which is equally good for discussing the on-shell properties such as conserved charges. Using
\be
d\widetilde\Psi=i_{\xi}\star \widetilde{F}
\ee
one finds that
\be
\delta H(g, A)-\delta H(g, \widetilde{A})=d(\Psi\delta A_{\1})-d(\widetilde{\Psi}\delta \widetilde{A}_{\1})\,.
\ee
Equivalently, we have
\be
\delta H(g,A)-d(\Psi\delta A_{\1})=\delta H(g,\widetilde{A})-d(\widetilde{\Psi}\delta \widetilde{A}_{\1})\,,
\ee
which means the improved infinitesimal Hamiltonian is invariant under electromagnetic duality.  We emphasize that although the Wald formalism has been applied to p-form systems such as \cite{Compere:2007vx}, the improvement needed to achieve the correct first law in the presence of magnetic charges has not been discussed.

In this paper, we report a novel case where the density of infinitesimal Hamiltonian needs a proper treatment in order for the first law to hold. Built upon our previous work \cite{Ma:2021opb}, we investigate the first law of thermodynamics for the dyonic strings carrying linear velocity in the context of $D=6$ IIA or heterotic string with leading $\a'$ corrections. In the IIA case, there exists a pair of mixed Chern-Simons (CS) terms of form
\be
-2B_{\2}\wedge {\rm tr}\big(R(\G_+)\wedge R(\G_+)\big)-2B_{\2}\wedge {\rm tr}\big(R(\G_-)\wedge R(\G_-)\big),
\ee
in which $\G_{\pm}$ refers to the torsionful connection with the torsion being $\pm H_{\3}$.
Of course, they can also be recast as
\be
2H_{\3}\wedge \mathrm{CS}_{\3}(\G_+)+2H_{\3}\wedge \mathrm{CS}_{\3}(\G_-)\,,
\ee
where $\mathrm{CS}_{\3}(\G_\pm)$ is the Chern-Simons form obeying $d\mathrm{CS}_{\3}(\G_\pm)= {\rm tr}(R(\G_\pm)\wedge R(\G_\pm))$. These two expressions differ by a total derivative term $d\big(2B_{\2}\wedge \mathrm{CS}_{\3}(\G_+)+2B_{\2}\wedge \mathrm{CS}_{\3}(\G_-)\big)$.

Usually if the Lagrangian is shifted by a total derivative, $\mathbf{L}\rightarrow \mathbf{L}+d\mathbf{\Lambda}$, the $\mathbf{\Theta}$ term also receives a shift
\bea
\mathbf{\Theta}\rightarrow\mathbf{\Theta}+\delta\mathbf{\Lambda}.\label{delta surface}
\eea
The Noether current $\mathbf{J}=\mathbf{\Theta}-i_\xi\mathbf{L}$ changes to
\bea
\mathbf{J}\rightarrow\mathbf{J}+\delta_{{\xi}}\mathbf{\Lambda}-i_\xi d\mathbf{\Lambda}.
\eea
If $\mathbf{\L}$ is a covariant quantity, $\delta_{{\xi}}\mathbf{\L}=\mathcal{L}_\xi\mathbf{\L}$, we then have
\bea
\mathbf{J}\rightarrow\mathbf{J}+di_\xi\mathbf{\Lambda}\,,
\eea
which means the Noether charge $\mathbf{Q}$ defined via $\mathbf{J}=d\mathbf{Q}$ acquires a shift according to
\bea
\mathbf{Q}\rightarrow\mathbf{Q}+i_\xi\mathbf{\Lambda}.\label{delta Noether charge}
\eea
Hence the $(D-2)$-form $\delta\mathbf{Q}[\xi]-i_{\xi}\mathbf{\Theta}$ would appear to be inert. However, since the total differential here is neither gauge invariant nor diffeomorphism invariant, it could have nontrivial effects on the density of infinitesimal Hamiltonian. The density of infinitesimal Hamiltonians resulting from the mixed
CS term $-2B_{\2}\wedge {\rm tr}(R(\G_\pm)\wedge R(\G_\pm))$ and $2H_{\3}\wedge \mathrm{CS}_{\3}(\G_\pm)$ were constructed in \cite{Bonora:2011gz}.\footnote{The construction of infinitesimal Hamiltonian for CS theories was revisited in \cite{Azeyanagi:2014sna}. However, both \cite{Bonora:2011gz} and \cite{Azeyanagi:2014sna} did not apply their formulas to study the entropy of 6D dyonic strings carrying the linear momentum and thus did not notice the infinitesimal Hamiltonian associated with the $-2B_{\2}\wedge {\rm tr}(R(\G_\pm)\wedge R(\G_\pm))$ term requires an improvement.} Using their results, we identify the possible closed but topologically nontrivial $4$-form that can be inserted into the naive result of $\delta\mathbf{Q}[\xi]-i_{\xi}\mathbf{\Theta}$ derived from the $-2B_{\2}\wedge {\rm tr}(R(\G_\pm)\wedge R(\G_\pm))$ term.

For dyonic string solutions carrying linear velocity, we show explicitly that this term takes different values at infinity and horizon. Thus its exclusion in $\delta\mathbf{Q}[\xi]-i_{\xi}\mathbf{\Theta}$ in literature leads to apparent violation of the first law. It should be noticed that when the solution is static or purely electric, this potential obstruction to the first law vanishes. This explains why in our previous work \cite{Ma:2021opb}, we had not noticed any problem with the first law derived using the $-2B_{\2}\wedge {\rm tr}(R(\G_\pm)\wedge R(\G_\pm))$ term. The first law derived from the $2H_{\3}\wedge \mathrm{CS}_{\3}(\G_\pm)$ works without any modification. This phenomenon may have to do with
the fact that general gauge CS terms are not globally defined on the base space of a principal bundle, while the gravitational CS terms are globally defined in spacetime due to the existence of a natural lift in the frame bundle, see for instance \cite{Bonora:1987bw}.

Once the first law is established, we can read off various thermodynamic quantities. To our surprise, the entropy that satisfies the first law can neither be obtained from the Wald formula applied to the $-2B_{\2}\wedge {\rm tr}(R(\G_\pm)\wedge R(\G_\pm))$ formulation
nor Tachikawa \cite{Tachikawa:2006sz} formula applied to the $2H_{\3}\wedge \mathrm{CS}_{\3}(\G_\pm)$ formulation. An immediate consequence of our
result is that upon taking the BPS limit, we obtain the entropy of the 3-charge BPS string solution in IIA string compactified on K3 that revises the previous results \cite{Castro:2007ci,Pang:2019qwq} obtained by directly applying the Wald-Tachikawa formula or attractor mechanism.\footnote{As shown by Sen, attractor mechanism is equivalent to Wald formula for extremal black holes.} Terms proportional to $\xi$ present in the infinitesimal Hamiltonian do no all vanish on the bifurcation horizon, some actually contribute. Also because the BPS string has zero temperature, one cannot verify the validity of its entropy using the first law of thermodynamics as we do here for black strings. The correct entropy for the 3-charge BPS string solution in IIA string compactified on K3 now has the desired property. It matches with the entropy of the 3-charge BPS string solution in heterotic string compactified on 4-torus upon performing the electromagnetic duality.

The outline of the paper is as follows.
In section \ref{section boost solution}, we give a brief review of the 3-charge dyonic string solution carrying the linear momentum in addition to the electric and magnetic charges, in 6D 2-derivative supergravity and its first law of thermodynamics. In section \ref{section type IIA}, we introduce the $\alpha'$ corrections arising from one loop terms in type IIA string compactified on K3. The $\a'$-corrected action contains terms of the form $-2B_{\2}\wedge {\rm tr}(R(\G_\pm)\wedge R(\G_\pm))$ which can also be written as $2H_{\3}\wedge \mathrm{CS}_{\3}(\G_\pm)$. We show how to obtain the correct first law for both formulations by improving the infinitesimal Hamiltonian in the $-2B_{\2}\wedge {\rm tr}(R(\G_\pm)\wedge R(\G_\pm))$ formulation with a topologically nontrival closed 4-form. We then compute the entropy of the $\a'$ corrected 3-charge solution using the first law, for which neither Wald formula nor Tachikawa formula provides the correct answer to the entropy. We also compute the Euclidean action and show it is compatible with the thermodynamic quantities computed from the infinitesimal Hamiltonian. In section \ref{section Heterotic}, we use IIA/heterotic duality to study the leading $\a'$ corrections to the thermodynamics of the 3-charge string solutions in 6D heterotic string compactified on 4-torus.  We conclude the paper in section 5.  In appendix A, we give the $\alpha'$-corrected perturbative solutions to the dyonic black string. In appendix B, we give the higher-derivative corrections to the infinitesimal Hamiltonian.

\section{3-charge dyonic black string in 6D 2-derivative supergravity}\label{section boost solution}
In this section, we review in detail how to interpret the first law of dyonic black string using the Wald procedure. The Lagrangian of $D=6$ minimal supergravity without higher derivative corrections is given by
\bea
\label{2dtheory}
e^{-1}\mathcal{L}_{\rm EH}=L\left(R+L^{-2}\nabla^\mu L\nabla_\mu L-\frac{1}{12}H_{\mu\nu\rho}H^{\mu\nu\rho}\right)\,,
\eea
where $e=\sqrt{-\det(g_{\mu\nu})}$, $L$ is the dilaton field, $H_{\mu\nu\rho}$ is a 3-form field strength of the 2-form potential $B_{\m\n}$, {\it i.e.}~$H_{\3}:=dB_{\2}$. Together, $(g_{\m\n},\,L,\,B_{\m\n})$ comprise the bosonic part of the 6D dilaton Weyl multiplet \cite{Bergshoeff:1985mz}. In this paper, we set 6D Newton's constant $G_6=1$. This theory admits a static black dyonic string solution \cite{Ma:2021opb} to which we can add the linear momentum by a Lorentz boost
\be
t\rightarrow c_3t+s_3x,\qquad x\rightarrow c_3x+s_3t,\qquad c_3\equiv\cosh{\delta_3},\qquad s_3\equiv\sinh{\delta_3}\,.
\label{boost}
\ee
The resulting solution takes the form
\bea
\label{non-BPS boost metric ansatz}
ds^2_6=&&D(r)\left(-h_1(r)dt^2+h_2(r)dx^2+2\omega(r) dtdx\right)+H_p(r)\left(\frac{dr^2}{f(r)}+r^2d\Omega^2_3\right),\cr
B_{\2}=&&2P\sqrt{1+\frac{\mu}{P}}\omega_{\2}+\sqrt{1+\frac{\mu}{Q_1}}A(r)dt\wedge dx\,,
\eea
where $\omega_{\2}=-\frac14\cos^2\frac{\theta}{2}d\phi\wedge d\chi$, so $d\omega_{\2}={\rm Vol}(S^3)$. Here the line element on $S^3$, $d\Omega^2_3=\frac{1}{4}(\sigma_3^2+d\Omega^2_2)$, is expressed as a $U(1)$ bundle over a $S^2$ in which $\sigma_3=d\chi-\cos{\theta}d\phi$, $d\Omega^2_2=d\theta^2+\sin^2{\theta}d\phi^2$.

After setting $s_3=-\sqrt{\frac{Q_2}{\mu}}$, the various functions in the solution are given by
\bea
L(r)=&&\frac{1}{D(r)H_p(r)},\quad  D(r)=A(r)=\frac{r^2}{r^2+Q_1},\quad H_p(r)=1+\frac{P}{r^2},\quad  f(r)=1-\frac{\mu}{r^2},\cr
h_1(r)=&&1-\frac{\mu}{r^2}-\frac{Q_2}{r^2}, \quad h_2(r)=1+\frac{Q_2}{r^2}, \quad  \omega(r)=-\frac{\sqrt{Q_2(\mu+Q_2)}}{r^2}.\label{Einstein solution}
\eea
The horizon of the black string is located at $r=r_h$ where the metric in the $dt\,,dx$ direction degenerates, i.e.
\be
(g_{tt}g_{xx}-g_{tx}^2)|_{r=r_h}=0\Rightarrow h_1(r_h)h_2(r_h)+\omega(r_h)^2=f(r_h)=0\,,
\label{horizon1}
\ee
from which we solve $r_h=\sqrt{\mu}$. The Killing vector $\xi$
\be
\xi=\partial_t+V_\mathrm{x}\partial_x,  \quad  V_{\rm x}=\sqrt{\frac{Q_2}{\mu +Q_2}}
\ee
becomes a null vector on the horizon i.e. $\xi^2|_{r=r_h}=0$. The linear momentum density along $x$-direction can be evaluated from the Komar integral
\be
P_\mathrm{x}=\frac{1}{16\pi}\int_{S^3}\star d\xi_x=\frac{ \pi}{4}  \sqrt{Q_2} \sqrt{\mu +Q_2},\quad \xi_x=\partial_{x}\,.
\label{momentum}
\ee
(Note that if $x$ is compact, $P_x$ can also interpreted as an angular momentum.) The temperature of the black string is defined through the surface gravity $\kappa$ using the Killing vector $\xi$
\be
\label{temp}
\kappa^2=-\frac{g^{\mu\nu}\partial_{\mu}\xi^2\partial_\nu\xi^2}{4\xi^2}\bigg|_{r=r_h},\quad T=\frac{\kappa}{2\pi}=\frac{1}{2\pi}\frac{\mu}{\sqrt{\mu +P} \sqrt{\mu +Q_1}\sqrt{\mu +Q_2}}\,.
\ee
The entropy density along the $x$-direction is computed using Iyer-Wald formula \cite{Iyer:1994ys}
\be
S=-\frac1{8}\int_{S^3} d\Omega_3\frac{\partial {\cal L}_{\rm EH}}{\sqrt{-g}\partial R_{\mu\nu\rho\sigma}}\epsilon_{\mu\nu}\epsilon_{\rho\sigma}\bigg|_{r=r_h}=\frac{1}{2}\pi ^2   \sqrt{\mu +P} \sqrt{\mu +Q_1}\sqrt{\mu +Q_2}\,,
\ee
where $\epsilon_{\mu\nu}$ is the binormal vector of the black string horizon satisfying $\epsilon_{\mu\nu}\epsilon^{\mu\nu}=-2$ and $d\O_3$ is the induced metric on the 3-sphere.
The electric and magnetic charges carried by the string are obtained as
\bea
Q_{\rm e}&=&\frac{1}{16\pi}\int_{S^3}L\star H_{\3}=\frac{ \pi}{4}  \sqrt{Q_1} \sqrt{\mu +Q_1}\,,\\
Q_{\rm m}&=&\frac{1}{16\pi}\int_{S^3}H_{\3}=\frac{ \pi}{4}  \sqrt{P} \sqrt{\mu +P}\,.
\eea
The corresponding electric and magnetic potential are computed from
\bea
\Phi_{\rm e}&=&\xi^\mu B_{\mu x}|_{r=\infty}-\xi^\mu B_{\mu x}|_{r=r_h}= \sqrt{\frac{Q_1}{\mu +Q_1}}\,,\\
\Phi_{\rm m}&=&\xi^\mu \widetilde{B}_{\mu x}|_{r=\infty}-\xi^\mu \widetilde{B}_{\mu x}|_{r=r_h}= \sqrt{\frac{P}{\mu +P}}\,,
\eea
where $\widetilde{B}_{\mu\nu}$ is the dual 2-form potential defined via $L\star H_{\3}=d\widetilde{B}_{\2}$. Similar to \cite{Ma:2021opb} the mass can be computed
using Brown-York surface Hamiltonian. The result is
\be
M=\frac{ 3\pi  }{8}\mu+\frac{ \pi  }{4}(Q_1+Q_2+P)\,.
\ee
One can check that the thermodynamic quantities satisfy the first law.
\be
dM=TdS+\Phi_{\rm e}dQ_{\rm e}+\Phi_{\rm m}dQ_{\rm m}+V_\mathrm{x}dP_\mathrm{x}.
\ee

Below we provide a different perspective based on the Wald procedure \cite{Iyer:1994ys}. For convenience, we fix the gauge $\xi^\mu B_{\mu x}|_{r=\infty}=0$ by shifting
$A(r)\rightarrow A(r)-1$. From the Lagrangian \eqref{2dtheory}, one first construct the conserved current \cite{Kim:2013zha}
\bea
\mathbf{J}_{\rm EH}&=&\frac1{5!}\epsilon_{\m\n\r\l\s\d}J^{\m}_{\rm EH}dx^\n\wedge dx^\r\wedge dx^\l\wedge dx^\s\wedge dx^\d\,,\\
J_{\rm{EH}}^{\mu}&=&\Theta^\m_{\rm EH}-\xi^{\mu}e^{-1}\mathcal{L}_{\mathrm{EH}}-2E^{\mu\nu}_{\rm{EH}}\xi_\nu
+2S^{\mu\lambda}_{\rm{EH}}B_{\nu\lambda}\xi^\nu\,,
\eea
in which
\bea
E_{\rm{EH}}^{\mu\nu}&=&LR^{\mu\nu}+L^{-1}(\nabla^\mu L)(\nabla^\nu L)-\frac{1}{2}g^{\mu\nu}e^{-1}\mathcal{L}_{\rm{EH}}-\frac{1}{4}LH^{2\mu\nu}+g^{\mu\nu}\Box L-\nabla^\mu\nabla^\nu L\,,\cr
S_{\rm{EH}}^{\mu\nu}&=&\frac{1}{2}\nabla_\rho\left(LH^{\rho\mu\nu}\right)\,,\quad \Theta^\m_{\rm EH}=\Theta_g^\mu+\Theta_B^\mu+\Theta_L^\mu, \cr
\Theta_{g}^\mu&=&Lg^{\mu\sigma}g^{\nu \rho}\left(\nabla_\rho\delta g_{\nu\sigma}-\nabla_\sigma\delta g_{\nu\rho}\right)+\delta g_{\nu\sigma}\left(g^{\nu\sigma}\nabla^\mu L-g^{\mu \nu}\nabla^\sigma L\right)\,,\cr
\Theta_B^\mu&=&-\frac{1}{2}LH^{\mu\nu\rho}\delta B_{\nu\rho}\,,\quad \Theta_L^\mu=2L^{-1}\nabla^\mu L\delta L\,.
\label{surface}
\eea
On-shell $d\mathbf{J}_{\rm EH}=0$ implies that $\mathbf{J}_{\rm EH}=d\mathbf{Q}_{\rm EH}$ where
\be
\mathbf{Q}_{\rm EH}=\frac1{4!2}\epsilon_{\a\b\m\n\r\l}Q^{\a\b}_{\rm EH}dx^\m\wedge dx^\n\wedge dx^\r\wedge dx^\l\,,\qquad Q_{\rm{EH}}^{\mu\nu}=Q_g^{\mu\nu}+Q_B^{\mu\nu}\,,
\ee
where
\be
Q_g^{\mu\nu}=-Lg^{\mu\sigma}g^{\nu\rho}\left(\nabla_\sigma\xi_\rho-
\nabla_\rho\xi_\sigma\right)-4\xi^{[\mu}\nabla^{\nu]}L, \quad Q_B^{\mu\nu}=-LH^{\mu\nu\rho}B_{\lambda\rho}\xi^\lambda\,.
\ee
Using the fact that $\xi$ is a Killing vector, when the perturbations $\delta g_{\m\n}\,,
\delta B_{\m\n}\,,\delta L$ obey the linearized field equations, one can show that
\be
d(\delta \mathbf{Q}_{\rm EH}-i_\xi \mathbf{\Theta}_{\rm EH})=0\,,
\ee
where $\mathbf{\Theta}_{\rm EH}=\frac1{5!}\epsilon_{\m\n\r\l\s\d}\Th^{\m}_{\rm EH}dx^\n\wedge dx^\r\wedge dx^\l\wedge dx^\s\wedge dx^\d$.

On substituting the details of the solution, we find that for $\xi=\partial_t+V_\mathrm{x}\partial_x$, $\int_{r=\infty\,,r_h}i_\xi \mathbf{\Theta}_L=0$
and
\bea
\int_{\infty}(\delta\mathbf{Q}_{B}[\xi]-i_{\xi}\mathbf{\Theta}_{B})&=&0\,, \quad \int_{r=r_h}(\delta\mathbf{Q}_{B}[\xi]-i_{\xi}\mathbf{\Theta}_{B})=\Phi_{\mathrm{e}}dQ_{\mathrm{e}}\,,\cr
\int_{\infty}(\delta\mathbf{Q}_{g}[\xi]-i_{\xi}\mathbf{\Theta}_{g})&=&dM-V_\mathrm{x}dP_\mathrm{x}, \quad\int_{r=r_h}(\delta\mathbf{Q}_{g}[\xi]-i_{\xi}\mathbf{\Theta}_{g})=TdS .
\eea
So apparently the infinitesimal Hamiltonian defined in \eqref{firstlaw} does not give rise to correct first law. This issue can be settled by improving the infinitesimal Hamiltonian density $\delta \mathbf{Q}_{\rm EH}-i_\xi \mathbf{\Theta}_{\rm EH}$ with the additional term
\be
-d(\Psi_{\1}\wedge\delta B_{\2})\,,\quad d\Psi_{\1}=i_\xi\star LH_{\3}\,.
\ee
For the solution \eqref{Einstein solution}, we have
\be
\Psi_{\1}=\frac{\sqrt{P(\mu +P)}}{P+r^2}(V_\mathrm{x}dt-dx)\,,
\ee
and
\bea
-\int_{\infty}d(\Psi_{\1}\wedge\delta B_{\2})=0\,,\quad-\int_{r=r_h}d(\Psi_{\1}\wedge\delta B_{\2})=\Phi_{\mathrm{m}}dQ_{\mathrm{m}}\,.
\eea
In fact, this can be seen more abstractly. Inclusion of the above total differential brings a term of the form $d(\Psi_{\1}\wedge\delta H_{\3})$ to the infinitesimal Hamiltonian.
Thus with the improvement, we have
\be
\delta{\cal H}_{\Sigma}=\int_{\Sigma}\left( \delta\mathbf{Q}[\xi]-i_{\xi}\mathbf{\Theta}-d(\Psi_{\1}\wedge\delta B_{\2})\right)\,.
\ee
the first law is indeed implied by
\be
\delta{\cal H}_{\infty}=\delta{\cal H}_{\mathcal{B}}\,.
\ee
It should be emphasized here however that despite the improvement, the Iyer-Wald formula for calculating the entropy is unchanged in this two-derivative theory. This story however no longer holds when we consider the $\alpha'$ corrections, discussed next.

\section{First law of 3-charge dyonic black string with mixed CS term}\label{section type IIA}

In this section, we extend the 2-derivative supergravity theory by supersymmetric Gauss-Bonnet term \cite{Novak:2017wqc,Butter:2018wss} and the Riemann tensor squared \cite{Bergshoeff:1986wc}
\bea
\label{action}
\mathcal{L}_{R+R^2}=&&\mathcal{L}_{\mathrm{EH}}+
\frac{\lambda_{\mathrm{GB}}}{16}\mathcal{L}_{\mathrm{GB}}+
\frac{\lambda_{\mathrm{Riem}^2}}{16}\mathcal{L}_{\mathrm{Riem}^2}\,,\\
\label{hdaction1}
e^{-1}\mathcal{L}_{\mathrm{GB}}=&&R_{\mu\nu\rho\sigma}R^{\mu\nu\rho\sigma}-
4R_{\mu\nu}R^{\mu\nu}+R^2+\frac{1}{6}RH^2
-R^{\mu\nu}H^2_{\mu\nu}+\frac{1}{2}R_{\mu\nu\rho\sigma}H^{\mu\nu\lambda}H^{\rho\sigma}_{\ \ \lambda}\cr
&&+\frac{5}{24}H^4+\frac{1}{144}\left(H^2\right)^2-\frac{1}{8}\left(H^2_{\mu\nu}\right)^2
+\frac{1}{4}\epsilon^{\mu\nu\rho\sigma\lambda\tau}B_{\mu\nu}R_{\rho\sigma\ \beta}^{\ \ \alpha}(\G_+)R_{\lambda\tau\ \alpha}^{\ \ \beta}(\G_+)\,,
\\
\label{hdaction2}
e^{-1}\mathcal{L}_{\mathrm{Riem}^2}=&&R_{\mu\nu\alpha\beta}(\G_-)R^{\mu\nu\alpha\beta}(\G_-)
+\frac{1}{4}\epsilon^{\mu\nu\rho\sigma\lambda\tau}B_{\mu\nu}R_{\rho\sigma\ \beta}^{\ \ \alpha}(\G_-)R_{\lambda\tau\ \alpha}^{\ \ \beta}(\G_-)
\,.
\eea
where in our convention $\sqrt{-g}\epsilon^{012345}=-1$. Here $R_{\mu\nu\ \beta}^{\ \ \alpha}(\G_{\pm})$ is the curvature defined with respect to the torsionful spin connection $\G^{\alpha}_{\pm\mu\beta}$
\bea
R_{\mu\nu\ \beta}^{\ \ \alpha}(\G_{\pm})=\partial_{\mu}\G^{\alpha}_{\pm\nu\beta}+
\G^{\alpha}_{\pm\mu\gamma}\G^{\gamma}_{\pm\nu\beta}-(\mu\leftrightarrow\nu)\,,
\quad \G^{\alpha}_{\pm\mu\beta}=\G^{\alpha}_{\mu\beta}\pm\frac{1}{2}H_{\mu\ \beta}^{\ \alpha}\,.
\eea
The shorthand notations for various contractions of $H_{\mu\nu\rho}$ are defined as
\bea
H^2=H_{\mu\nu\rho}H^{\mu\nu\rho}\,,\quad H^2_{\mu\nu}=H_{\mu\rho\sigma}H_{\nu}^{\ \rho\sigma}\,,\quad H^4=H_{\mu\nu\sigma}H_{\rho\lambda}^{\ \ \sigma}H^{\mu\rho\delta}H^{\nu\lambda}_{\ \ \ \delta}\,.
\eea
The combination with $\lambda_{\mathrm{GB}}=\lambda_{\mathrm{Riem}^2}=\alpha'$ describes the leading $\a'$ correction to
 the NS-NS sector of IIA compactified on K3 \cite{Liu:2013dna} and thus is compatible with 6D (1,1) supersymmetry.

After including the leading $\alpha'$-correction, all the functions will be perturbed
\bea
\label{non BPS perturbation}
L(r)&=&L_0(r)+\delta L(r)\,,\quad D(r)=D_0(r)+\delta D(r)\,,\quad A(r)=D_0(r)+\delta A(r)\,,\cr
h_1(r)&=&h_{1,0}(r)+\delta h_1(r),\ \ \ h_2(r)=h_{2,0}(r)+\delta h_2(r),\ \ \ \omega(r)=\omega_0(r)+\delta\omega(r)\,,\cr
f(r)&=&f_0(r)+\delta f(r)\,,
\eea
where the subscript ``0" labels the 2-derivative solution \eqref{Einstein solution}.
Again, the $\a'$ corrected 3-charge dyonic solution can be obtained from the $\a'$ corrected 2-charge solution by the Lorentz boost \eqref{boost}. Functions $\delta L(r)$, $\delta D(r)$, $\delta A(r)$, $\delta f(r)$ remain the same as the unboosted solution, while there
is a mixing among the metric components in the $dt\,,dx$ direction
\be
\delta h_1(r)=c_3^2\delta h-2c_3s_3\delta\widetilde{\omega}\,,\quad
\delta h_2(r)=-s_3^2\delta h+2c_3s_3\delta\widetilde{\omega}\,,\quad
\delta \omega(r)=-c_3s_3\delta h+(c_3^2+s_3^2)\delta\widetilde{\omega}\,,
\ee
where $s_3=-\sqrt{\frac{Q_2}{\mu}}$. All the perturbed functions $\delta L(r)$, $\delta D(r)$, $\delta A(r)$, $\delta f(r)$, $\delta h$, $\delta\widetilde{\omega}$ can be found in Appendix \ref{app solution}. The horizon defined by (\ref{horizon1}) receives correction too
\bea
r_h\rightarrow \sqrt{\mu}+\alpha'\delta r\,,\quad\delta r=\frac{\sqrt{\mu } \left(\mu  (\mu +P)-4 \mu  Q_1-4 Q_1^2\right)}{16 Q_1 (\mu +Q_1)^2}
-\frac{\mu ^{5/2} (\mu +P)}{16 Q_1^2 (\mu +Q_1)^2}\log \left(1+\frac{Q_1}{\mu }\right)\,.
\eea
Up to ${\cal O}(\a')$, the Killing vector which becomes null on the horizon is still given by
\be
\xi=\partial_t+V_\mathrm{x}\partial_x,  \quad  V_{\rm x}=\sqrt{\frac{Q_2}{\mu +Q_2}}\,,
\ee
using which we can obtain the temperature according to \eqref{temp}
\be
T=\frac{1}{2\pi}\frac{\mu}{\sqrt{\mu +P} \sqrt{\mu +Q_1}\sqrt{\mu +Q_2}}-\frac{\mu Q_1 (5 \mu +4 Q_1) }{4 \pi  \sqrt{\mu +P} (\mu +Q_1)^{5/2} (\mu +2 Q_1)\sqrt{\mu +Q_2}}\alpha'\,.
\ee
The electric and magnetic charges are computed by the standard way
\bea
Q_{\rm e}&=&\frac{1}{16\pi}\int_{S^3}\star \mathcal{M}_{\3}=\frac{1}{4} \pi  \sqrt{Q_1} \sqrt{\mu +Q_1}\,,\\
Q_{\rm m}&=&\frac{1}{16\pi}\int_{S^3}H_{\3}=\frac{1}{4} \pi  \sqrt{P} \sqrt{\mu +P}\,.
\eea
where $d\star \mathcal{M}_{\3}=0$ is the $B_{\2}$ field equation with
\bea
&&{{-6}}\mathcal{M}^{\mu\nu\rho}=LH^{\mu\nu\rho}+\frac{\alpha'}{16}
\Big(12R^{\lambda[\mu}H_{\lambda}{}^{\nu\rho]}-2RH^{\mu\nu\rho}-\frac16H^2H^{\mu\nu\rho}
-2H^{[\mu}{}_{\lambda\sigma}H^{2\nu|\lambda|,\rho]\sigma}+4\square H^{\mu\nu\rho}\cr
&&\qquad+6R_{\alpha\beta}^{\ \ [\mu\nu}(\G_+)\star H^{\rho]\alpha\beta}+2\star\mathrm{CS}^{\mu\nu\rho}(\G_+)-6R_{\alpha\beta}^{\ \ [\mu\nu}(\G_-)\star H^{\rho]\alpha\beta}+2\star\mathrm{CS}^{\mu\nu\rho}(\G_-)
\Big)\,.
\eea
The electric and magnetic potential are given by
\bea
\Phi_{\rm e}&=&\xi^\mu B_{\mu x}|_{r=\infty}-\xi^\mu B_{\mu x}|_{r=r_h}=\sqrt{\frac{Q_1}{\mu +Q_1}}+\frac{\mu  \sqrt{Q_1} (5 \mu +4 Q_1) }{2 (\mu +Q_1)^{5/2} (\mu +2 Q_1)}\alpha'\,,\\
\Phi_{\rm m}&=&\xi^\mu \widetilde{B}_{\mu x}|_{r=\infty}-\xi^\mu \widetilde{B}_{\mu x}|_{r=r_h}=\sqrt{\frac{P}{\mu +P}}.
\eea

Below we compute the mass and the linear momentum by integrating the infinitesimal Hamiltonian associated with Killing vectors $\partial_t$ and $\partial_x$ respectively. Now the charge $\mathbf{Q}$ and the surface term $\mathbf{\Th}$ both receive higher derivative corrections. Their expressions can be found in Appendix \ref{app wald}. The mass can be read off from
the gravitational contribution to the infinitesimal Hamiltonian associated with $\partial_t$, namely,
\bea
\delta M&=&\delta \mathcal{H}_\infty[\partial_t]=\int_{\infty}( \delta\mathbf{Q}[\partial_t]-i_{\partial_t}\mathbf{\Th})\nn\\
\Rightarrow
M&=&\frac{ 3\pi  }{8}\mu+\frac{ \pi  }{4}(Q_1+Q_2+P)-\frac{3 \pi  \mu ^2 (3 \mu +2 Q_1) }{32 (\mu +Q_1)^2 (\mu +2 Q_1)}\alpha'\,.
\eea
Similarly, we can obtain the momentum by replacing $\partial_t$ with $\partial_x$
\bea
\delta P_\mathrm{x}&=&-\delta \mathcal{H}_\infty[\partial_x]=-\int_{\infty}( \delta\mathbf{Q}[\partial_x]-i_{\partial_x}\mathbf{\Theta})\nn\\
&\Rightarrow& P_\mathrm{x}=\frac{ \pi}{4}  \sqrt{Q_2} \sqrt{\mu +Q_2}\,.
\eea
It appears that the linear momentum does not receive $\a'$ correction.

In order to compute the entropy, we now investigate the complete infinitesimal Hamiltonian
associated with $\xi=\partial_t+V_{\rm x}\partial_x$. The higher derivative corrections to the
infinitesimal Hamiltonian are given in Appendix \ref{app wald}. In the presence of the non-diffeomorphism invariant CS term, the infinitesimal Hamiltonian in general takes the form
\be
\delta{\cal H}_{\Sigma}=\int_{\Sigma}( \delta\mathbf{Q}[\xi]-i_{\xi}\mathbf{\Theta}[\d\phi]-\mathbf{\Sigma}[\xi])\,.
\ee
As showed in the previous section, the above infinitesimal Hamiltonian missed the magnetic contribution to the first law and should be
improved by adding $-d(\Psi_{\1}\wedge\delta B_{\2})$
\bea
\delta \mathcal{H}_\Sigma[\xi]=\int_{\Sigma}( \delta\mathbf{Q}[\xi]-i_{\xi}\mathbf{\Theta}[\xi]-d(\Psi_{\1}\wedge\delta B_{\2}))\,,
\label{magnetic H}
\eea
where the 1-form $\Psi_{\1}$ is defined by
\be
d\Psi_{\1}=i_\xi\star \mathcal{M}_{\3}\,.
\ee
For the $\a'$ corrected solution, it takes the form
\be
\Psi_{\1}=\frac{\sqrt{P} \sqrt{\mu +P}}{P+r^2}\Upsilon(V_\mathrm{x}dt-dx)\,,
\ee
where $\Upsilon$ is
\bea
\Upsilon&=&1+\frac{\alpha'}{8 Q_1^2 (P+r^2) (\mu +Q_1)^2 (Q_1+r^2)^2}\Big[\cr
&&Q_1\Big(\mu ^2 r^2 (P+r^2)(2 r^2-\mu )+4 \mu  Q_1^3 (2 \mu -5 r^2)+ \mu ^2Q_1^2 (4 \mu +P-15 r^2)\cr
&&+4Q_1^4 ( \mu -2 r^2) +\mu ^2Q_1 (-\mu  P+3 P r^2-5 \mu  r^2+3 r^4)
\Big)\cr
&&-\mu ^2 (P+r^2) (Q_1+r^2)^2 (2 r^2-\mu )\log (1+\frac{Q_1}{r^2})
\Big]+\mathcal{O}(\alpha'^2).
\eea
At this moment, we need to distinguish the infinitesimal Hamiltonians associated with
two formulations of the higher derivative actions. In the original actions \eqref{hdaction1} and \eqref{hdaction2}, the mixed CS term is given by $-2B_{\2}\wedge {\rm tr}(R(\G_\pm)\wedge R(\G_\pm))$. We denote the corresponding total infinitesimal Hamiltonian as
$\delta \mathcal{H}^{(1)}_\S$. By adding a total differential $d(2B_{\2}\wedge \mathrm{CS}_{\3}(\G_\pm))$, one obtains another formulation in terms of $2H_{\3}\wedge \mathrm{CS}_{\3}(\G_\pm)$. We denote the corresponding total infinitesimal Hamiltonian as
$\delta \mathcal{H}^{(2)}_\S$. Here these two infinitesimal Hamiltonians have been improved with the addition of the term $-d(\Psi_{\1}\wedge\delta B_{\2})$.

We now evaluate the infinitesimal Hamiltonian at spatial infinity and horizon using the 3-charge string solution. It turns out that we indeed have
\be
\delta \mathcal{H}^{(2)}_\infty=\delta \mathcal{H}^{(2)}_{\mathcal{B}}\,,
\ee
from which we can read off the entropy that satisfies the first law automatically
\be
\label{entropy non BPS type A}
S_{\rm IIA}=\frac{1}{2} \pi ^2  \sqrt{\mu +P} \sqrt{\mu +Q_1}\sqrt{\mu +Q_2}+\frac{\pi ^2 Q_1  \sqrt{\mu +P}\sqrt{\mu+Q_2 } (5 \mu +4 Q_1) }{4 (\mu +Q_1)^{3/2} (\mu +2 Q_1)}\alpha'\,,
\ee
which reproduces the result in \cite{Ma:2021opb} when $Q_2=0$. However, for $\delta \mathcal{H}^{(1)}_\S$, apparently, its value at the infinity is not equal to its value at the horizon.
This means there is a topological obstruction forbidding us to apply the Gauss theorem.
After comparing the infinitesimal Hamiltonians associated with two different formulations of the action, we find that the density of $\delta\mathcal{H}^{(1)}_\Sigma$ needs a further improvement by adding a term of the form
$\frac{\a'}{16}d\Pi$ where
\bea
\Pi&=&\delta(2B_{\2}\wedge\G^a_{+ b}i_\xi\G^b_{+ a}+2B_{\2}\wedge\G^a_{- b}i_\xi\G^b_{- a})\cr
&&+i_\xi (2B_{\2}\wedge\G^a_{+ b}\wedge\delta\G^b_{+ a}+2B_{\2}\wedge\G^a_{- b}\wedge\delta\G^b_{- a}).
\eea
With the second improvement, we find indeed
\be
\delta \mathcal{H}^{(1)}_\infty=\delta \mathcal{H}^{(1)}_{\mathcal{B}}\,.
\ee
From this we can read off the same entropy \eqref{entropy non BPS type A}. By contrast, the Wald entropy formula would lead to different entropies of the black strings for these two different formulation of the theory.

Some remarks need to make here. The entropy formula \eqref{entropy non BPS type A} does not coincide with one computed using Wald formula applied to the original formulation of the action with the $-2B_{\2}\wedge {\rm tr}(R(\G_\pm)\wedge R(\G_\pm))$ term. Also it cannot be obtained using the Tachikawa formula applied to the second formulation with $2H_{\3}\wedge \mathrm{CS}_{\3}(\G_\pm)$ term. To be specific, we have
\be
S_{\rm W}=S_{\rm IIA}-\frac{\pi ^2\sqrt{PQ_2}}{4 \sqrt{\mu +Q_1}} \alpha'\,,\quad
S_{\rm T}=S_{\rm IIA}+\frac{\pi ^2\sqrt{PQ_2}}{4 \sqrt{\mu +Q_1}}\alpha'\,,
\label{wrong type A entropy}
\ee
from which we see that only when $PQ_2=0$, the Wald-Tachikawa formula yields the right answer in agreement with our previous calculation \cite{Ma:2021opb}. We also note that in the extremal case $\m=0$, the entropy obtained from Tachikawa formula
equals the one obtained from Sen's approach. We present the computation based on Sen's approach in Appendix C.

For the 3-charge solution, the improvement term takes the form
\bea
d\Pi_\infty=0\,,\quad d\Pi_{r=r_h}&=&-\frac{\pi\mu}{\sqrt{P+\mu}\sqrt{Q_2+\mu}(Q_1+\mu)}\Big(
\frac{\mu  \sqrt{Q_2}}{\sqrt{P} (\mu +P)}\delta P-\frac{2 \sqrt{P} \sqrt{Q_2}}{\mu +Q_1}\delta Q_1\cr
&&+\frac{2 \sqrt{P}}{\sqrt{Q_2}}\delta Q_2+\frac{\sqrt{P} \sqrt{Q_2} (3 \mu +2 P+Q_1)}{(\mu +P) (\mu +Q_1)}\delta\mu
\Big)\,.
\eea
We see that it vanishes on $P=0$, $Q_2=0$ or $\m=0$. This explains why we had not encountered it in our previous work dealing with the 2-charge string solution without linear momentum.

We have checked that the near horizon geometry for the non-extremal black string is perfectly
smooth describing $\mathbb{R}\ltimes {\rm Rindler}_2\times S^3$. The reason that Wald or Tachikawa formula does not apply to our case is due to the fact on the horizon not only terms proportional $\nabla_{a}\xi_{b}$ contribute, but also terms proportional to Killing vector $\xi$ have nonvanishing contributions. After a careful calculation we find that in the improved infinitesimal Hamiltonian evaluated on the horizon $\delta{\cal H}_{\cal B}$, terms proportional to $\delta B$ contribute to $\Phi_{\mathrm{e}}dQ_{\mathrm{e}}+\Phi_{\mathrm{m}}dQ_{\mathrm{m}}$ which is known previously. However, terms proportional to $\xi$, $\delta g,\,\delta L$  which one would naively think to vanish on the horizon also contribute. We denote these contributions as $\delta{\cal H}_{\cal B}[\xi,\,\delta g,\,\delta L]$.  Together with the variation of the Wald charge term, they yield $TdS$ term in the first law. As a concrete example, in the formulation of the 4-derivative action with $-2B_{\2}\wedge {\rm tr}(R(\G_\pm)\wedge R(\G_\pm))$, we obtain
\bea
\delta{\cal H}_{\cal B}[\xi,\,\delta g,\,\delta L]&=&\frac{\mu (P-Q_1)}{4 (\mu +P) (\mu +Q_1)}\delta\mu\cr
&&+\frac{\alpha '}{32 Q_1^3 (\mu +P) (\mu +Q_1)^4}\big(A_1\delta\mu+A_2\delta Q_1+A_3\delta Q_2+A_4\delta P\,,
\big)
\eea
where
\bea
&&A_1=-\frac{Q_1^2}{(\mu +P)^{3/2} (\mu +2 Q_1) \sqrt{\mu +Q_2}}\Big(2  Q_1\sqrt{P Q_2} (\mu +P) (\mu +Q_1)^2 (\mu +2 Q_1) \cr
&&\times\big(\mu  (6 \mu +Q_1)-P (Q_1-4 \mu )\big)+\mu  \sqrt{\mu +P} \sqrt{\mu +Q_2}\Big(\mu ^5 (P-13 Q_1)\cr
&&+\mu ^4 (P-7 Q_1)(2 P+Q_1)+\mu ^2 Q_1^2 (46 P^2+133 P Q_1-29 Q_1^2)\cr
&&+12 P Q_1^4 (5 P-2 Q_1)+2 \mu  Q_1^2 (59 P^2 Q_1-2 P^3+19 P Q_1^2-13 Q_1^3)\cr
&&+\mu ^3 (-P^2 Q_1+P^3+48 P Q_1^2+6 Q_1^3)
\Big)
\Big)\,,\cr
&&A_2=-\mu ^2 Q_1\Big(\mu  P^2 (2 \mu +3 Q_1)+P (2 \mu ^3+\mu ^2 Q_1-7 \mu  Q_1^2-4 Q_1^3)\cr
&&-Q_1 (2 \mu ^3+3 \mu ^2 Q_1-4 \mu  Q_1^2-4 Q_1^3)\Big)+2 \mu ^4 (\mu +P) (P-Q_1) (\mu +2 Q_1) \log (\frac{\mu +Q_1}{\mu })\,,\cr
&&A_3=\frac{2 \mu  \sqrt{P} Q_1^4 \sqrt{\mu +P} (\mu +Q_1)^2}{\sqrt{Q_2} \sqrt{\mu +Q_2}}\,,\cr
&&A_4=\mu ^3 Q_1 (P-Q_1) (\mu +Q_1) \Big(Q_1-\mu  \log (\frac{\mu +Q_1}{\mu })\Big)\,.
\eea
In the BPS limit $\mu\rightarrow0$, the entropy density and various charges become
\bea
S&=&\frac{\pi ^2}{2}  \sqrt{PQ_1Q_2} +\frac{\pi ^2}{2}\sqrt{\frac{PQ_2}{Q_1}}\alpha'=\frac{\pi ^2}{2}  \sqrt{P(Q_1+2\alpha')Q_2}+{\cal O}(\a')^2\,,\cr
Q_{\rm e}&=&\frac{\pi}4Q_1\,,\quad Q_{\rm m}=\frac{\pi}4P\,,\quad P_{\rm x}=\frac{\pi}4Q_2\,,\quad
M=\frac{ \pi  }{4}(Q_1+Q_2+P)\,.
\label{BPSS0}
\eea
The near horizon limit of the BPS solution is given by an extremal BTZ $\times S^3$ whose entropy is the same as the entropy of the full BPS string solution, since the entropy is determined by the geometry of the horizon. Thus from the entropy of the extremal BTZ black hole, one can read off one of the central charges in the dual 2D CFT. In string unit, $\ell_s=1$ $G_6=\frac{\pi^2}2$, the entropy density can be expressed as\footnote{The factor of $2\pi$ in the entropy fomula is recovered had we compactified the $x$-direction with period 2$\pi$ and calculated the entropy instead of its density.}
\be
\label{BPSS}
S=\sqrt{P(Q_1+2)Q_2}\,,
\ee
where parameters $P,\,Q_1,\,Q_2$ are integer valued corresponding the number of NS5 compactified on K3, number of fundamental string and excitation level of momentum modes respectively. The formula \eqref{BPSS} implies that one of central charge in the 2D CFT dual to IIA string on $AdS_3\times S^3\times$ K3 is given by $c=6P(Q_1+2)$. The finite shift in $Q_1$ is reminiscent of the central charge in the CFT dual to heterotic string in the $AdS_3\times S^3\times T^4$ background \cite{Kutasov:1998zh}. In fact, since IIA string on K3 is dual to heterotic string on 4-torus, it is natural to expect that the CFT duals of two scenarios should have the same central charge. As we will see, this is indeed the case in the next section.

We end this section by computing the Euclidean action of the $\a'$-corrected 3-charge string solution. Similar to the 2-charge string solution, we evaluate the Euclidean action using the background substraction method. The total action is thus given by
\bea
\label{Euclidean action}
I_{E}=I_0+I_{\rm hd}+I_{\mathrm{GHY}}-I_{c}\,,
\eea
where $I_0$ and $I_{\rm hd}$ are the leading and subleading bulk actions, while $I_{\mathrm{GHY}}$ and $I_c$ are the string frame analogue of Gibbons-Hawking-York boundary term and background subtraction term given below
\bea
I_{\mathrm{GHY}}=\frac1{8\pi}\int_{\partial M} d^5x\sqrt{-h}LK,\quad I_c=\frac1{8\pi}\int_{\partial M} d^5x\sqrt{-h}L\bar{K}\,.
\eea
Here we emphasize that Euclidean action seems to be insensitive to the choice of $-2B_{\2}\wedge {\rm tr}(R(\G_\pm)\wedge R(\G_\pm))$ or $2H_{\3}\wedge \mathrm{CS}_{\3}(\G_\pm)$. Both formulations yield the same answer. The reason is that the contribution from $-2B_{\2}\wedge {\rm tr}(R(\G_+)\wedge R(\G_+)) (2H_{\3}\wedge \mathrm{CS}_{\3}(\G_+))$ cancels with that from $-2B_{\2}\wedge {\rm tr}(R(\G_-)\wedge R(\G_-)) (2H_{\3}\wedge \mathrm{CS}_{\3}(\G_-) )$.

In the intermediate steps, the boundary is located at some large value of the radial coordinate $r=r_c$ which will be taken to infinity eventually. $K$ is the trace of the extrinsic curvature of the $r=r_c$ hypersurface embedded in $\a'$-corrected string solution. $\bar{K}$ is the trace of the extrinsic curvature of the $r=r_c$ hypersurface embedded in the flat background metric below
\bea
d\bar{s}^2_6=D(r_c)(h_1(r_c)d\tau^2+h_2(r_c)dx^2)+dR^2+R^2d\Omega^2_3,\quad R^2=r^2H_p(r)\,.
\eea
After substituting the solution in (\ref{Euclidean action}), we obtain
\bea
I_E=\beta G\,,\quad \beta=T^{-1}\,,\quad G=\frac{\pi}8(\mu+2P)-\frac{\pi  \mu  (9 \mu +8 Q_1) }{32 (\mu +Q_1)^2}\alpha'\,,\label{free energy}
\eea
which indeed satisfies
\bea
G=M-TS-\Phi_{\mathrm{e}}Q_{\mathrm{e}}-V_{\mathrm{x}}P_{\mathrm{x}}\,.
\eea

Instead we can also use Reall-Santos method \cite{Reall:2019sah}. A consequence of \cite{Reall:2019sah} is that at fixed conserved charges, the entropy is given by $-I_{\rm hd}$ evaluated on the leading order solution. For the 3-charge string solution, we have
\be
I_{\rm hd}=-\frac{\pi \b \mu  (9 \mu +8 Q_1) }{32 (\mu +Q_1)^2}\alpha'=-\frac{\pi ^2  \sqrt{\mu +P}\sqrt{\mu+Q_2 } (9 \mu +8 Q_1)}{16 (\mu +Q_1)^{3/2}}\alpha'\,,
\ee
which does match with $\Delta S(M,\,Q_{\rm e},\,Q_{\rm m})$.

\section{Six-dimensional Heterotic/Type \uppercase\expandafter{\romannumeral2}A duality}\label{section Heterotic}

In this section, we use the IIA/heterotic duality to study the leading $\a'$ corrections to the thermodynamics of the 3-charge string solutions in 6D heterotic string compactified on 4-torus. In the heterotic string, the leading $\a'$ corrections arise at the tree level \cite{Bergshoeff:1985mz,Bergshoeff:1989de }. When compactified on 4-torus, the heterotic string is dual to IIA string compactified on K3 \cite{Hull:1994ys}. After discarding the Yang-Mills field, the bosonic Lagrangian takes the form
\bea
\label{het theory}
\mathcal{L}=L\left(R+L^{-2}\nabla^\mu L\nabla_\mu L-\frac{1}{12}\widetilde{H}_{\mu\nu\rho}\widetilde{H}^{\mu\nu\rho}
+\frac{\alpha'}{8}R_{\mu\nu\alpha\beta}(\G_+)R^{\mu\nu\alpha\beta}(\G_+)\right),
\eea
where the 3-form field strength
\bea
\widetilde{H}_{\3}=H_{\3}+\frac{1}{4}\alpha'\mathrm{CS}_{\3}(\G_+)
\eea
satisfies a non-trivial Bianchi identity
\bea
\label{non-trivial Bianchi}
d\widetilde{H}_{\3}=\frac{1}{4}\alpha'R^a_{\ b}(\G_+)\wedge R^b_{\ a}(\G_+).
\eea
The equations of motion in linear order of $\alpha'$ are
\bea
0&=&R+L^{-2}(\nabla L)^2-2L^{-1}\Box L-\frac{1}{12}\widetilde{H}^2+\frac{1}{8}\alpha' R_{\mu\nu\alpha\beta}(\G_+)R^{\mu\nu\alpha\beta}(\G_+)\,,\cr
0&=&R_{\mu\nu}-L^{-1}\nabla_\mu\nabla_\nu L+L^{-2}(\nabla_\mu L)(\nabla_\nu L)-\frac{1}{4}\widetilde{H}_{\mu\nu}^2+\frac{1}{4}\alpha'R_{\mu \alpha\beta\gamma}(\G_+)R_\nu^{\ \alpha\beta\gamma}(\G_+)\,,\cr
0&=&d\left(L\star \widetilde{H}_{\3}\right).\label{het form EOM}
\eea
Up to this $\a'$ order, the above field equations are mapped to those in IIA case via \cite{Liu:2013dna}
\be
\label{map}
L^{\rm \sst{IIA}}\star H_{\3}^{\rm IIA}=\widetilde{H}_{\3}^{\rm het},\quad L^{\rm IIA}g_{\mu\nu}^{\rm IIA}=g_{\mu\nu}^{\rm het},\quad L^{\rm IIA}=\frac{1}{{L^{\rm het}}}\,.
\ee
We have checked that the 3-charge solution in IIA indeed maps to a 3-charge string solution in heterotic side. Similar to the 2-charge case studied in \cite{Ma:2021opb}, one needs to perform a shift on the $B_{\2}$
\be
B_{\2}\rightarrow B_{\2}-\Lambda_{\2},\quad \Lambda_{\2}=\frac{8P}{r^2+P}\sqrt{1+\frac{\mu}{P}}\omega_{\2}
\ee
so that the ansatz \eqref{non-BPS boost metric ansatz} is still applicable. As pointed out in \cite{Ma:2021opb}, that for the 2-charge solution, the thermodynamic quantities in the heterotic side is related to those in the IIA side by the change of variables
\be
P\rightarrow Q_1,\qquad Q_1\rightarrow P.
\ee
Now since the map \eqref{map} commutes with the Lorentz boost, we can obtain the thermodynamic quantities for the heterotic string. The consereved charges can also be computed independently using other methods except for the entropy which can only be derived from the first law. The results are listed below
\bea
M^{(\mathrm{het})}&=&\frac{ 3\pi  }{8}\mu+\frac{ \pi  }{4}(Q_1+P+Q_2)-\frac{3 \pi  \mu ^2 (3 \mu +2 P) }{32 (\mu +P)^2 (\mu +2 P)}\alpha '\,,\cr
T^{(\mathrm{het})}&=&\frac{1}{2\pi}\frac{\mu}{\sqrt{\mu +P} \sqrt{\mu +Q_1}\sqrt{\mu +Q_2}}-\frac{\mu P (5 \mu +4 P) }{4 \pi  \sqrt{\mu +Q_1} \sqrt{\mu +Q_2}(\mu +P)^{5/2} (\mu +2 P)}\alpha '\,,\cr
S^{(\mathrm{het})}&=&\frac{1}{2} \pi ^2  \sqrt{\mu +P} \sqrt{\mu +Q_1}\sqrt{\mu +Q_2}+\frac{\pi ^2 P  \sqrt{\mu +Q_1}\sqrt{\mu+Q_2 } (5 \mu +4 P) }{4 (\mu +P)^{3/2} (\mu +2 P)}\alpha'\,,\cr
Q_{\rm e}^{(\mathrm{het})}&=&\frac{\pi}{4}  \sqrt{Q_1} \sqrt{\mu +Q_1},\quad \Phi_{\rm e}^{(\mathrm{het})}=\sqrt{\frac{Q_1}{\mu +Q_1}}\,,\cr
Q_{\rm m}^{(\mathrm{het})}&=&\frac{ \pi}{4}  \sqrt{P} \sqrt{\mu +P},\quad \Phi_{\rm m}^{(\mathrm{het})}=\sqrt{\frac{P}{\mu +P}}+\frac{\mu  \sqrt{P}  (5 \mu +4 P)}{2 (\mu +P)^{5/2} (\mu +2 P)}\alpha ',\cr
P_{\rm x}^{(\mathrm{het})}&=&\frac{ \pi}{4}  \sqrt{Q_2} \sqrt{\mu +Q_2},\quad V_{\rm x}^{(\mathrm{het})}=\sqrt{\frac{Q_2}{\mu +Q_2}}\label{het non BPS }
\eea
We find that the Wald-Tachikawa entropy formula and the recently proposed  covariantized Wald-Tachikawa entropy \cite{Elgood:2020nls} yield
\be
S_{\rm WT}=S^{\rm (het)}+\frac{\pi ^2\sqrt{Q_1Q_2}}{4 \sqrt{\mu +P}}\alpha'\,,
\ee
which does not satisfy the first law of thermodynamics. In the BPS limit $\mu\rightarrow0$, the entropy becomes to
\be
S^{(\mathrm{het})}=\frac{\pi ^2}{2}  \sqrt{PQ_1Q_2} +\frac{\pi ^2}{2}\sqrt{\frac{Q_1Q_2}{P}}\alpha'=\frac{\pi ^2}{2}  \sqrt{(P+2\alpha')Q_1Q_2}.
\ee

\section{Conclusions}

In this work, using the Wald procedure, we performed a careful study on the leading $\a'$ corrections to the first law of thermodynamics for 3-charge string solutions in 6D supergravity arising from IIA string compactified on K3.  The low energy effective action contains mixed CS terms of the form $-2B_{\2}\wedge {\rm tr}(R(\G_\pm)\wedge R(\G_\pm))$ which can also be recast as $2H_{\3}\wedge \mathrm{CS}_{\3}(\G_\pm)$ up to a total derivative. We found that the infinitesimal Hamiltonian derived from the former formulation does not lead to the desired equality $\delta{\cal H}_{\infty}=\delta{\cal H}_{\mathcal{B}}$ that implies the first law of thermodynamics. Thus it cannot be used to define the entropy consistently. On the other hand, the infinitesimal Hamiltonian derived from the second formulation involving $2H_{\3}\wedge \mathrm{CS}_{\3}(\G_\pm)$ gives straightly the desired equality $\delta{\cal H}_{\infty}=\delta{\cal H}_{\mathcal{B}}$. By comparing the infinitesimal Hamiltonians associated with the 2 formulations, we realized that the density of the infinitesimal Hamiltonian associated with the first formulation can be improved by adding a closed but non-exact 4-form (65,67) whose value at spatial infinity and horizon is just right to restore the first law. With this improvement, both formulations yield infinitesimal Hamiltonians
obeying the equality $\delta{\cal H}_{\infty}=\delta{\cal H}_{\mathcal{B}}$. From the first law of thermodynamics we read off the $\a'$-corrected entropy of the 3-charge black string solution. Taking the extremal limit, we obtained the $\a'$-corrected entropy for BPS 3-charge string solution \eqref{BPSS0} that is different from the previous results \cite{Castro:2007ci} obtained by directly applying the Wald-Tachikawa formula or attractor mechanism. We also reproduced our entropy result using another method proposed by Reall and Santos \cite{Reall:2019sah}. Thus we have found a case where the Wald-Tachikawa formula does
not apply and developed a procedure to find the entropy consistent with the first law.

To our surprise, the entropy cannot be obtained directly from either Wald formula or Tachikawa formula.  Our results illustrate a danger of using either formulae to compute directly the entropy of extremal black holes, where the first-law of black hole thermodynamics cannot provide a consistent check at zero temperature. In our computation, we noticed that on the horizon terms proportional to $\xi$ can also contribute to the first law instead of being vanishing. One possible reason is that we usually define the linearized solution by
\be
\delta(c_i) \phi=\delta c_i\partial_{c_i}\phi(c_i)
\ee
where $\phi$ is  a shorthand notation for all the fields in the theory and $c_i$ denotes the physical parameters. This seems to be a natural way to generate the linearized solutions. However, it is likely that for generic $c_i$, the linearized solutions obtained this way do not satisfy the smoothness condition needed in the abstract proof given by
\cite{Wald:1999wa,Kay:1988mu,Jiang:2021pna} where terms proportional to $\xi$ all vanish on the horizon.

Our way of improving the infinitesimal Hamiltonian can be readily generalized to other
mixed CS terms in diverse dimensions, for instance the $A\wedge R\wedge R$ and $A\wedge F\wedge F$ terms. These terms appear in 5D 4-derivative supergravity actions together with the curvature squared terms \cite{Ozkan:2013nwa} and are relevant in the precision test of AdS$_5$/CFT$_4$ correspondence \cite{CHLS,Bobev1,Bobev2,Liu:2022sew}. Compared to the Euclidean action method \cite{Reall:2019sah}, the improved infinitesimal Hamiltonian is applicable to all solutions regardless of its asymptotic structure since it is insensitive to the asymptotics of the spacetime metric. Finally, it would be interesting to apply our procedure to study first law
of thermodynamics for other black objects such as black rings in models with Chern-Simons
interactions \cite{Rogatko:2006hck}.

\section*{Acknowledgement}

Liang Ma thanks Xing-Hui Feng, Hai-Shan Liu and Run-Qiu Yang for discussions.
The work is supported in part by the National Natural Science Foundation of China (NSFC) grants No.~11875200, No.~11935009 and No.~12175164.

\appendix

\section{Perturbative solutions in IIA}\label{app solution}

In this appendix section, we present the perturbations to the 2-charge string solution in the type IIA case which are used to generate the perturbative solutions to the 3-charge string solutions. We first define
\bea
C_0&=&2 C_1 Q_1+C_3 Q_1 (\mu +2 Q_1)-2 \left(C_4+C_7\right) Q_1 (\mu +Q_1)+2 C_5 (\mu +Q_1)\,,\nn\\
X(r)&=&(1+\frac{Q_1}{\mu })^{-2}\left((3+\frac{2Q_1}{\mu })\log (1-\frac{\mu }{r^2})-\frac{\mu^2}{Q_1^2}\log (1+\frac{Q_1}{r^2})\right).
\eea
Solutions to various perturbations are given by
\bea
&&\delta\widetilde{\omega}=\frac{\mu  \sqrt{P} \sqrt{\mu +P} }{2 r^2 (P+r^2) (Q_1+r^2)}\a'\,,\nn\w2
\label{equ coupling f 8}
&&\delta f=\frac{C_1 }{r^2}(1+\frac{P}{r^2})-\frac{\mu  \a'}{8 Q_1 r^4}(1+\frac{Q_1}{\mu })^{-1}\Big(\mu +2 Q_1-\frac{Q_1 (5 \mu +4 Q_1)-P (\mu +2 Q_1)}{r^2}\cr
&&-\frac{Q_1 (\mu  P+4 Q_1^2+4 \mu  Q_1)}{r^4}\Big)(1+\frac{Q_1}{r^2})^{-2}+\frac{C_2}{2 \mu  r^2}(1+\frac{P}{r^2})\log (1-\frac{\mu }{r^2})-\frac{\a'}{8r^2}(1+\frac{P}{r^2})X(r),\nn\w2
&&\delta h=\frac{C_1}{\mu  r^2}(\mu +2 P-\frac{\mu  P}{r^2})+\frac{C_2}{\mu  r^2}(1+\frac{P}{r^2})+C_7 (1-\frac{\mu }{r^2})-\frac{\a'}{8 Q_1 r^2}(1+\frac{Q_1}{\mu })^{-1}\Big(2 (\mu +2 Q_1)\cr
&&+\frac{-\mu ^2+4 P (\mu +2 Q_1)+4 Q_1^2+2 \mu  Q_1}{r^2}+\frac{2 P^2 (\mu +2 Q_1)-2 \mu  P (\mu +2 Q_1)-4 \mu  Q_1 (\mu +Q_1)}{r^4}\cr
&&+\frac{P }{r^6}\Big(P (-\mu ^2+4 Q_1^2+2 \mu  Q_1)+4 \mu  Q_1 (\mu +Q_1)\Big)\Big)(1+\frac{Q_1}{r^2})^{-1}(1+\frac{P}{r^2})^{-1}\cr
&&+\frac{1}{2 \mu ^3}(1+\frac{Q_1}{\mu })^{-1}\Big( (C_0 \mu +4 C_1 P Q_1)(1-\frac{\mu }{r^2})-C_2\Big(\mu -2 Q_1-\frac{2 P (\mu +Q_1)+\mu  (2 \mu -Q_1)}{r^2}\cr
&&+\frac{\mu  P (\mu +Q_1)}{r^4}\Big)
\Big)\log (1-\frac{\mu }{r^2})-\frac{ \a'}{8 \mu }(2-\frac{\mu }{r^2})(1+\frac{P}{r^2})X(r),\nn\w2
&&\delta D=-\Big(\frac{C_0}{2 \mu  r^2}+\frac{ C_1 P Q_1}{\mu ^2 r^2}(2+\frac{\mu }{r^2})+\frac{C_2 Q_1}{\mu ^2 r^2}(2-\frac{P}{r^2})-\frac{C_3 Q_1}{2 r^2}-\frac{C_4}{2}(1+\frac{Q_1}{r^2})\Big)(1+\frac{Q_1}{r^2})^{-2}\cr
&&+\frac{\a'}{8 \mu  Q_1 r^2}(1+\frac{Q_1}{\mu })^{-1}\Big((\mu +2 Q_1) (3 \mu +4 Q_1)+\frac{4 Q_1^2}{r^2} \Big((P+4 Q_1)+\mu ^2 (3 P+7 Q_1)\cr
&&+4 \mu  Q_1 (2 P+5 Q_1)\Big) +\frac{2 Q_1}{r^4} \Big(2 P (\mu +Q_1) (\mu +2 Q_1)-\mu  Q_1 (2 \mu +3 Q_1)-P^2 (\mu +2 Q_1)\Big)\cr
&&+\frac{P Q_1 }{r^6}\Big(\mu ^2 (P-11 Q_1)-4 \mu  Q_1 (P+4 Q_1)-4 Q_1^2 (2 P+Q_1)\Big)\cr
&&-\frac{P Q_1^2 }{r^8}\Big(P (4 Q_1^2+2 \mu  Q_1-\mu ^2)+4 \mu  Q_1 (\mu +Q_1)\Big)
\Big)(1+\frac{Q_1}{r^2})^{-4}(1+\frac{P}{r^2})^{-1}\cr
&&-\frac{1}{2 \mu ^3}\Big(C_0 \mu +4 C_1 P Q_1+C_2 Q_1(2-\frac{\mu }{r^2})(2-\frac{P}{r^2}
\Big)(1+\frac{Q_1}{r^2})^{-2}\log (1-\frac{\mu }{r^2})\cr
&&+\frac{\a'}{8 \mu^2}\Big(3 \mu +4 Q_1-\frac{2 Q_1 (\mu +P)}{r^2}+\frac{\mu  P Q_1}{r^4}\Big)(1+\frac{Q_1}{r^2})^{-2}X(r),\nn\w2
&&\delta A=C_6-\Big(\frac{C_0 Q_1}{2 \mu ^2 r^2}(1+\frac{Q_1}{\mu })^{-1}(1-\frac{\mu }{r^2})
+\frac{C_1 P Q_1}{\mu ^3 r^2}(1+\frac{Q_1}{\mu })^{-1}(2Q_1+\frac{\mu  (\mu -Q_1)}{r^2})\cr
&&+\frac{C_2 Q_1}{2 \mu ^3 r^2}(1+\frac{Q_1}{\mu })^{-1}(\mu+4 Q_1 -\frac{2 P (\mu +Q_1)+3 \mu  Q_1}{r^2})+\frac{C_3 Q_1^2}{2 r^4}+\frac{C_5}{2 r^2}(1+\frac{Q_1}{r^2})\Big)(1+\frac{Q_1}{r^2})^{-2}\cr
&&+\frac{\a'}{8 \mu  r^2}(1+\frac{Q_1}{\mu })^{-1}\Big(4 (\mu +2 Q_1)-\frac{5 \mu ^2+2 P (\mu +2 Q_1)-16 Q_1^2-2 \mu  Q_1}{r^2}\cr
&&-\frac{1}{r^4}\Big(P (-\mu ^2+8 Q_1^2+4 \mu  Q_1)+5 \mu  Q_1 (3 \mu +4 Q_1)
\Big) -\frac{Q_1 }{r^6}\Big(P (-\mu ^2+4 Q_1^2+2 \mu  Q_1)\cr
&&+2 \mu  Q_1 (\mu +Q_1)\Big) \Big)(1+\frac{Q_1}{r^2})^{-4}-\frac{Q_1}{8 \mu ^4}(1+\frac{Q_1}{\mu })^{-1}\Big(4 C_0 \mu  (1-\frac{\mu }{r^2})+16 C_1 P Q_1 (1-\frac{\mu }{r^2})\cr
&&+4 C_2\Big(\mu +4 Q_1-\frac{2 P (\mu +Q_1)+\mu  (2 \mu +5 Q_1)}{r^2}+\frac{\mu  P (\mu +Q_1)}{r^4}
\Big)\Big)(1+\frac{Q_1}{r^2})^{-2}\log(1-\frac{\mu }{r^2})\cr
&&+\frac{Q_1\a'}{8 \mu^2}(4-\frac{5 \mu +2 P}{r^2}+\frac{\mu  P}{r^4})(1+\frac{Q_1}{r^2})^{-2}X(r),\nn\w2
\label{equ coupling L 8}
&&\delta L=\Big(\frac{C_0 \mu +2 C_1 P (2 Q_1-\mu )+2 C_2 (P+2 Q_1)-C_3 \mu ^2Q_1}{2 \mu ^2 r^2}
+C_8(1+\frac{Q_1}{r^2})\Big)(1+\frac{P}{r^2})^{-1}\cr
&&-\frac{\a'}{8 \mu  Q_1 r^2}(1+\frac{Q_1}{\mu })^{-1}\Bigg((\mu +2 Q_1) (3 \mu +2 P+4 Q_1)
-\frac{1}{r^2}\Big(-8 Q_1^2 (P+2 Q_1)+\mu ^2 (P-7 Q_1)\cr
&&-4 \mu  Q_1 (P+5 Q_1)\Big)-\frac{Q_1 }{r^4}\Big(P (\mu ^2-4 Q_1^2-2 \mu  Q_1)+2 \mu  Q_1 (2 \mu +3 Q_1)\Big)\Bigg)(1+\frac{Q_1}{r^2})^{-2}(1+\frac{P}{r^2})^{-1}\cr
&&+\frac{1}{8 \mu ^4}(1+\frac{Q_1}{\mu })^{-1}\Bigg(
2 (C_0 \mu +4 C_1 P Q_1)(\mu +2 Q_1-\frac{\mu  Q_1}{r^2})+2 C_2\Big(
3 \mu ^2+4 P (\mu +Q_1)+8 Q_1^2\cr
&&+8 \mu  Q_1-\mu\frac{ 2 P (\mu +Q_1)+Q_1 (\mu +4 Q_1) }{r^2}\Big)\Bigg)(1+\frac{P}{r^2})^{-1}\log (1-\frac{\mu }{r^2})\cr
&&-\frac{\a'}{8 \mu^2}\Big(
3 \mu +2 P+4 Q_1-\frac{\mu  (P+2 Q_1)}{r^2}\Big)(1+\frac{P}{r^2})^{-1}X(r)\,.
\eea
The physical solution, with an appropriate horizon and asymptotic falloffs, corresponds to the parameter choice
\bea
\label{solution 1}
&&\{C_1,\,C_2,\,C_3,\,C_4,\,C_5,\,C_6,\,C_7,\,C_8\}\cr
&=&\{0,\frac{\mu ^2 (3 \mu +2 Q_1)\a' }{4 (\mu +Q_1)^2},\frac{3 \mu ^2 (3 \mu +2 Q_1) \a'}{4 Q_1 (\mu +Q_1)^2 (\mu +2 Q_1)}, 0,0,0,0,0\}\,.
\eea
\section{$\a'$ corrections to infinitesimal Hamiltonian}\label{app wald}

In this appendix section, we present the contributions to the Noether charge $\mathbf{Q}$ and surface term $\mathbf{\Th}$ from the higher derivative action. In the presence of non-diffeomorphism invariant action such as
the $2H_{\3}\wedge \mathrm{CS}_{\3}(\G_\pm)$ term, there is a third contribution
to the infinitesimal Hamiltonian denoted by $\mathbf{\Sigma}[\xi]$ \cite{Bonora:2011gz}. Thus the most general form
of the infinitesimal Hamiltonian in a theory with CS type interaction is given by
\be
\delta{\cal H}_{\Sigma}=\int_{\Sigma}( \delta\mathbf{Q}[\xi]-i_{\xi}\mathbf{\Theta}[\d\phi]-\mathbf{\Sigma}[\xi])\,.
\ee

Due to the CS terms, the Noether charge and surface term contain covariant part and non-covariant part
\bea
Q^{\mu\nu}=Q_{\mathrm{cov}}^{\mu\nu}+Q_\mathrm{{nc}}^{\mu\nu},\quad \Theta^{\mu}=\Theta_{\mathrm{cov}}^{\mu}+\Theta_\mathrm{{nc}}^{\mu}.
\eea
The covariant parts in $\mathbf{Q}$ and $\mathbf{\Th}$ consist of several pieces
\bea
Q^{\mu\nu}_{\mathrm{cov}}=&&Q^{\mu\nu}_{\mathrm{cov}1}+
\frac{\lambda_{\mathrm{GB}}}{16}Q^{\mu\nu}_{\mathrm{cov}2+}
+\frac{\lambda_{\mathrm{Riem}^2}}{16}Q^{\alpha\beta}_{\mathrm{cov}2-},\\
\Theta^\mu_{\mathrm{cov}}=&&\Theta^\mu_{\mathrm{cov}1}+
\frac{\lambda_{\mathrm{GB}}}{16}\Theta^\mu_{\mathrm{cov}2+}
+\frac{\lambda_{\mathrm{Riem}^2}}{16}\Theta^\mu_{\mathrm{cov}2-}\,.
\eea
in which
\bea
Q^{\mu\nu}_{\mathrm{cov}1}&=&-2P^{\mu\nu\gamma\delta}\nabla_\gamma\xi_\delta+
4\xi_\delta\nabla_\gamma
P^{\mu\nu\gamma\delta}+6\mathcal{M}^{\mu\nu\rho}B_{\gamma\rho}\xi^\gamma\,,\cr
&&+\frac{\lambda_{\mathrm{Riem}^2}}{16}\Big(-2\nabla^{[\mu}H_{\alpha\beta\gamma}
H^{\nu]\beta\gamma}\xi^\alpha+2\nabla^{[\mu}H^{\nu]\beta\gamma}
H_{\alpha\beta\gamma}\xi^\alpha-2\nabla_\alpha H^{[\mu}_{\ \ {{\beta\gamma}}}H^{\nu]\beta\gamma}\xi^\alpha\Big)\,,\\
\Theta^\mu_{\mathrm{cov}1}&=&2P^{\mu\beta\gamma\delta}\nabla_\delta\delta g_{\beta\gamma}-2\delta g_{\beta\gamma}\nabla_\delta P^{\mu\beta\gamma\delta}+3\mathcal{M}^{\mu\beta\gamma}\delta B_{\beta\gamma}\cr
&&+\frac{\lambda_{\mathrm{Riem}^2}}{16}\left[\Big((\nabla^\alpha H^{\beta\nu\rho})H^\mu_{\ \nu\rho}-(\nabla^\mu H^{\beta\nu\rho})H^\alpha_{\ \nu\rho}-(\nabla^\beta H^{\mu\nu\rho})H^\alpha_{\ \nu\rho}
\Big)\delta g_{\alpha\beta}\right.\nn\\
&& \left.\qquad +\frac{2}{3}(\nabla^\mu H^{\alpha\beta\gamma})\delta H_{\alpha\beta\gamma}\right]\,,
\eea
with
\bea
P_{\mu\nu\rho\sigma}&=&\frac{\lambda_{\mathrm{GB}}}{16}\Big(
2R_{\mu\nu\rho\sigma}+2\left(g_{\mu\sigma}R_{\nu\rho}-
g_{\mu\rho}R_{\nu\sigma}+g_{\nu\rho}R_{\mu\sigma}-g_{\nu \sigma}R_{\mu\rho}\right)+\left(g_{\mu\rho}g_{\nu\sigma}-g_{\mu\sigma}g_{\nu\rho}\right)R\cr
&&+\frac{1}{2}H^2_{\mu\nu,\rho\sigma}+\frac{1}{4}\left(g_{\mu \sigma}H^2_{(\nu\rho)}-g_{\nu\sigma}H^2_{(\mu\rho)}+
g_{\nu\rho}H^2_{(\mu\sigma)}-g_{\mu\rho}H^2_{(\nu\sigma)}\right)\cr
&&+\frac{1}{12}\left(g_{\mu\rho}g_{\nu\sigma}-g_{\mu\sigma}g_{\nu\rho}\right)H^2\Big)
+\frac{\lambda_{\mathrm{Riem}^2}}{16}\left(2R_{\mu\nu\rho\sigma}-
\frac{1}{2}H^2_{\mu\nu,\rho\sigma}\right)\,,\\
\mathcal{M}^{\alpha\beta\gamma}&=&\frac{\lambda_{\mathrm{GB}}}{16}\Big(
R^{\lambda\tau[\alpha\beta}H^{\gamma]}_{\ \ \lambda\tau}
-2R^{\lambda[\alpha}H_{\lambda}^{\ \beta\gamma]}+\frac{1}{3}RH^{\alpha\beta\gamma}+\frac{1}{36}H^2H^{\alpha\beta\gamma}\cr
&&-\frac{1}{2}H^{2\lambda[\alpha}H_{\lambda}^{\ \beta\gamma]}+\frac{5}{6}H^{[\alpha}_{\ \ \lambda\tau}H^{2\beta|\lambda|,\gamma]\tau}\Big)+\frac{\lambda_{\mathrm{Riem}^2}}{16}\Big(
-\frac{2}{3}\square H^{\alpha\beta\gamma}\cr
&&-R^{\lambda\tau[\alpha\beta}H^{\gamma]}_{\ \ \lambda\tau}+\frac{1}{2}H^{2\lambda[\alpha}H_{\lambda}^{\ \beta\gamma]}-\frac{1}{2}H^{[\alpha}_{\ \ \lambda\tau}H^{2\beta|\lambda|,\gamma]\tau}\Big)\,,
\eea
where $H^{2\alpha\beta,\lambda\tau}=H^{\alpha\beta\delta}H^{\lambda\tau}_{\ \ \delta}$. The second covariant part come from
the mixed CS action and is universal for $-2B_{\2}\wedge {\rm tr}(R(\G_-)\wedge R(\G_-))$ or $2H_{\3}\wedge \mathrm{CS}_{\3}(\G_\pm)$
\bea
Q^{\alpha\beta}_{\mathrm{cov}2\pm}&=&2R_{\mu\nu}^{\ \ \alpha\beta}(\G_\pm)\star H^{\mu\nu\rho}\xi_\rho-4R_{\mu\nu}^{\ \ \rho[\alpha}(\G_\pm)\star H^{\beta]\mu\nu}\xi_\rho\pm 6R_{\mu\nu}^{\ \ [\beta\rho}(\G_\pm)\star H^{\alpha]\mu\nu}B_{\rho\sigma}\xi^\sigma\,,\\
\Theta^\mu_{\mathrm{cov}2\pm}&=&2R_{\alpha\beta}^{\ \ \ \mu\nu}(\G_\pm)\star H^{\gamma\alpha\beta}\delta g_{\nu\gamma}\mp3R_{\alpha\beta}^{\ \ \ [\mu\nu}(\G_\pm)\star H^{\rho]\alpha\beta}\delta B_{\nu\rho}\,.
\eea
The $-2B_{\2}\wedge {\rm tr}(R(\G_-)\wedge R(\G_-))$ or $2H_{\3}\wedge \mathrm{CS}_{\3}(\G_\pm)$ term contributes to  the noncovariant parts differently. For $-2B_{\2}\wedge {\rm tr}(R(\G_+)\wedge R(\G_+))$ or $-2B_{\2}\wedge {\rm tr}(R(\G_-)\wedge R(\G_-))$ term, one obtains \cite{Bonora:2011gz}
\bea
\mathbf{Q}^{\rm nc}_{\rm{BRR}\pm}[\xi]&=&-4B_{\2}\wedge R^a_{\ b}(\G_\pm)\Lambda^b_{\ a}-4B_{\2}\wedge R^a_{\ b}(\G_\pm)i_\xi\G^b_{\pm a}\,, \cr
\mathbf{\Theta}^{\rm nc}_{\rm{BRR}\pm}&=&-4B_{\2}\wedge R^a_{\ b}(\G_\pm)\wedge\delta\G^b_{\pm a}\,,
\eea
where $\Lambda^b_{\ a}=\partial_a\xi^b$. Here due to the noncovariant nature, it is more convenient to present these term in differential form. For $2H_{\3}\wedge \mathrm{CS}_{\3}(\G_+)+2H_{\3}\wedge \mathrm{CS}_{\3}(\G_-)$ term, one obtains \cite{Bonora:2011gz}
\bea
\mathbf{Q}^{\rm nc}_{\mathrm{HCS}\pm}[\xi]=&&4H_{\3}\wedge \G^a_{\pm b}\Lambda^b_{\ a}+2H_{\3}\wedge\G^b_{\pm a} i_\xi\G^a_{\pm b}-2\mathrm{CS}_{\3}(\G_{\pm})\wedge i_\xi B_{\2}\,,\cr
\mathbf{\Theta}^{\rm nc}_{\mathrm{HCS}\pm}=&&2H_{\3}\wedge\G^a_{\pm b}\wedge\delta\G^b_{\pm a}+2\mathrm{CS}_{\3}(\G_{\pm})\wedge\delta B_{\2}\,.
\eea
Although the difference between $-2B_{\2}\wedge {\rm tr}(R(\G_\pm)\wedge R(\G_\pm))$ and $2H_{\3}\wedge \mathrm{CS}_{\3}(\G_\pm)$
is just $d(2B_{\2}\wedge\mathrm{CS}_{\3}(\G_\pm))$, the difference between the Noether charge and the surface term is more than
what's been discussed in the introduction for the covariant total derivative term (\ref{delta surface},\ref{delta Noether charge}). Instead, we find
\bea
&&\mathbf{Q}^{\rm nc}_{\mathrm{BRR}\pm}[\xi]+i_\xi(2B_{\2}\wedge\mathrm{CS}_{\3}(\G_\pm))+
d\Pi_{1\pm}=\mathbf{Q}^{\rm nc}_{\mathrm{HCS}\pm}[\xi]\,,\cr
&&\mathbf{\Theta}^{\rm nc}_{\mathrm{BRR}\pm}+\delta(2B_{\2}\wedge\mathrm{CS}_{\3}(\G_\pm))+d\Pi_{2\pm}
=\mathbf{\Theta}^{\rm nc}_{\mathrm{HCS}\pm}\,,
\label{transformation rule 2}
\eea
where locally one can write
\be
\Pi_{1\pm}=2B_{\2}\wedge\G^a_{\pm b}i_\xi\G^b_{\pm a}\,,\quad
\Pi_{2\pm}=2B_{\2}\wedge\G^a_{\pm b}\wedge\delta\G^b_{\pm a}\,.
\ee
In our derivation of \eqref{transformation rule 2}, we have used the fact that $\xi$ is Killing vector and in the coordinate system adopted in the computation, its components are constant. Thus
\bea
\mathcal{L}_\xi\G^a_{\pm b}=0,\quad \Lambda^b_{\ a}=0.
\eea
Using the fact that ${\cal L}_{\xi}\Pi_{2\pm}=0$ on shell, we obtain
\be
\delta\mathbf{Q}^{\rm nc}_{\mathrm{HCS}\pm}[\xi]-i_{\xi}\mathbf{\Theta}^{\rm nc}_{\mathrm{HCS}\pm}=
\delta\mathbf{Q}^{\rm nc}_{\mathrm{BRR}\pm}[\xi]-i_{\xi}\mathbf{\Theta}^{\rm nc}_{\mathrm{BRR}\pm}+d\delta\Pi_{1\pm}+di_{\xi}\Pi_{2\pm}\,.
\ee

Finally, we give the expression of last noncovariant term arising only from $2H_{\3}\wedge \mathrm{CS}_{\3}(\G_\pm)$
\be
\mathbf{\Sigma}_{\mathrm{HCS}\pm}=2\delta B_{\2}\wedge  \G^a_{\pm b}\wedge d\L^b_{\ a}\,.
\ee

\section{Entropy of the extremal black string obtained from Sen's approach}
In this section, we demonstrate
that Sen' approach to the computation of the entropy of the extremal black string
yields the same answer as the Tachikawa formula for the case of IIA string compactified on K3.
We first study the near horizon geometry of the
extremal black string corresponding to $\m=0$ in the black string solution, as the of $\a'$-corrected temperature Eq.(52) goes to 0. According to Eq.(50), the horizon is located at $r=0$
and thus the near horizon limit is achieved by zooming in the region near $r=0$.
We define the new variables
\bea
\rho&=&\frac{r^2}{Q'_1}\,,\quad a^2=\frac{Q'_2}{Q'_1}\,,\quad t'=\frac{2t}{a\sqrt{P}}\,,\nn\\
Q'_1&=&Q_1+\a'\,,\quad  Q'_2=Q_2(1-2\frac{\a'}{Q_1})\,,
\eea
in terms of which, the near horizon geometry becomes
\bea
ds^2_{\rm NH}&=&\frac{P}4(-\r^2dt'^2+\frac{d\r^2}{\r^2})+a^2(dx+\frac{\sqrt{P}}{2a}\r dt')^2+Pd\Omega_3^2\,,\nn\\
H_{(3)}&=&P{\rm Vol(S^3)}+ \frac{a\sqrt{P}}{2} dt'\wedge dx\wedge d\rho\,,\quad L=\frac{Q'_1}{P}\,,
\eea
which is U(1)$\ltimes AdS_2\times S^3$ or extremal BTZ$\times S^3$.

To construct the entropy function, we then make the ansatz based on the near horizon geometry
\bea
ds^2_{\rm NH}&=&\frac{\ell^2}4(-\r^2dt'^2+\frac{d\r^2}{\r^2})+(\frac{\ell}{2e_2})^2(dx+e_2\r dt')^2+Pd\Omega_3^2\,,\nn\\
H_{(3)}&=&P{\rm Vol(S^3)}+ e_1 dt'\wedge dx\wedge d\rho\,,\quad L=L_h\,,
\eea
where $\ell,\,e_1,\,e_2,\,L_h$ are constants to be determined from extremizing the entropy function.
Then we can write down the entropy density function using Eq.(3.17) given in \cite{Pang:2019qwq}
by setting $\l_1=\l_2=\a'$. If we do not compactify $x$ to have $2\pi$ period,
the entropy density function is given by
\bea
&&{\cal E}=e_1q_1+e_2q_2-f(\ell,\,e_1,\,e_2,\,L_h)\,,\nn\\
&&f(\ell,\,e_1,\,e_2,\,L_h)=\frac{P^{\ft32}}2\Big[L_h(\frac{4e_2e_1^2}{\ell^3}+\frac{\ell^3}{2e_2P}-\frac{3\ell}{4e_2})
+\a'(\frac{24e_2^3e_1^4}{\ell^9}-\frac{3e_2e_1^2}{\ell^5}\nn\\
&&\qquad\qquad \qquad\qquad\qquad -\frac{4e_2e_1^2}{\ell^3P}-\frac{3e_1}{2\ell^2\sqrt{P}}+\frac{3}{32e_2\ell}-\frac{1}{2e_1\sqrt{P}}-\frac{3\ell}{4e_2P})\Big]\,,
\eea
where $q_1,\,q_2$ are conjugate variables of $e_1,\,e_2$ and we have set $1/G_3=4P^{\ft32}$ according to the convention of \cite{Pang:2019qwq}. Extremizing ${\cal E}$
with respect to $\ell,\,e_1,\,e_2,\,L_h$ leads to
\be
\ell=\sqrt{P}\,,\quad e_2=\frac12\sqrt{\frac{P(L_hP+2\a')}{q_2}}\,,\quad L_h=\frac{q_1+\a'}P\,,\quad 4e_1e_2=P\,.
\label{sol1}
\ee
Comparing solutions above for $\ell,\,e_1,\,e_2,\,L_h$ to those arising from the near horizon geometry of the
asymptotically flat extremal black string, we find that
\be
q_1=Q_1\,,\quad q_2=Q_2\,,
\label{sol2}
\ee
which is reasonable since $Q_1,\,Q_2$ are the conserved charges carried by the black string even with $\a'$ corrections.
Finally, after substituting \eqref{sol1} and \eqref{sol2} to the entropy density function,
we obtain
\be
{\cal E}=\sqrt{PQ_2(Q_1+3\a')}\,.
\ee
To see this matched with the one obtained from Tachikawa formula, we take the second formula in
Eq.(69) and set $\m=0$. Using Eq.(66), we see that
the result obtained from Tachikawa formula coincides with the one derived from Sen's approach after recovering
the $G_6$ dependence in the denominator and using the fact that in string unit $\ell_s = 1,\, G_6 =\ft{\pi^2}2$.

\end{document}